\documentclass[12pt]{article}
\usepackage{epsfig}
\usepackage[dvips]{color}

\begin{document}
\title{General Chaotic Behaviors of Heavy ion Collisions at
Intermediate Energy Based on Dynamical Transport Model}

\author{\small  Yong-Zhong Xing$^{1}$ \footnote{E-mail:
 yzxing@tsnu.edu.cn}, Wen-Xia Wang$^{2}$, Hong-Fei Zhang $^{1,2}$ and Yu-Ming Zheng$^{1,3}$ }
\date{}
\maketitle

\begin{center}
$^{1}${\small Institute for the Fundamental physics, Tianshui Normal
University, Gansu, Tianshui 741000, P. R. China}\\
$^{2}${ School of Nuclear Science and Technology, Lanzhou University, Lanzhou 730000, P. R. China}\\
$^{3}${\small China Institute of Atomic Energy, P.O. Box 275(10),
Beijing  102413, P. R. China}\\
\end{center}

\baselineskip 0.3in.
\begin{center}{\bf Abstract}\end{center}
\hskip 0.2in
\begin{small}
\begin{quote}

Motivated to explore the dynamical mechanism of multi-fragmentation phase transition and expand our knowledge on the deterministic chaos in nonlinear dynamic process, we study systematically the nonlinear dynamical behaviors of heavy ion collisions at intermediate energy. In order to highlight the general characteristics of the collision dynamics, in the present paper, we simulate a simple collision system, $Ca+Ca$, and obverse the performances of the typical nonlinear characteristic quantities such as the production of generalized entropy and the fractal structure of the semiclassical phase space during the reaction process. The multifragmentation entropy, information dimension and the dynamical fluctuations of fragment mass distribution in the final state of the reaction are also evaluated. Our results about the intermediate processes and the final products of the reaction are verified mutually so that confirm strongly the existence of nonlinear chaos in the heavy ion collisions at intermediate energy. Meanwhile, we find that the typical characteristics of the critical phase transition obtained in the analysis of the final production of the reaction are relatively more clear in the vicinity of the incident energies $E=100 MeV$.

\end{quote}
\end{small}


{\bf PACS} number(s): 24.60.Lz, 25.70.Pq

\newpage
\baselineskip 0.3in

{\section{Introduction}}

The dynamical mechanism of the multifragmentation, which is an essential phenomenon occurring in the Heavy Ion Collision(HIC) at intermediate energies, has been explored for a long time and such studies  are still current focus[1-5]. Correspondingly, spinodal instability, liquid-gas phase transition, critical behavior, power law, etc, associated with the process of the HIC have been investigated intensively and explained variously in the last decades[6-11]. In fact, such phenomena link intimately to the non-linear dynamics or deterministic chaos which became an independent science in the 70's of the last century[12-16]. Chaotic dynamics has opened up a new way for us to understand the complex world so that it has been considered as a bridge between dynamics and statistics. Specifically, Chaos is a randomness produced by deterministic dynamics which is essentially different from statistical randomness. In the theoretic studies of the heavy-ion nuclear reactions at intermediate energy the multifragmentation can be described very well by utilizing  either the  dynamical transport model or the statistics model, separately. This consistency  shows naturally light on how to understand the role of deterministic chaotic dynamics playing in bridging the dynamics and the statistics.  Therefore, to study the special manifestation of the non-linear dynamical processes of the HICs and/or the non-statistical fluctuation in the final states of the reactions make sense for us not only to understand the collision dynamics of HICs more comprehensively but also to enrich our knowledge of the deterministic chaos.

Actually, so far, a series of physical quantities reflecting the chaotic characteristics of nonlinear dynamics are available for us to describe the complex system or chaos. Are these quantities suitable for the description of heavy ion collision dynamics though it is essentially a high nonlinear dynamical process?  what is the common performances of the chaotic characteristics in the heavy ion collisions at intermediate energy? Such questions have not yet been fully answered. Although there are many physical quantities such as that quoted above can reflect the nonlinear dynamic behaviour of nuclear reactions, some of them are usually obtained in the analysis of different experiments in the help of various theoretic models, e.g., percolation and/or cascade model, and some are even found in the investigation of ideal models, such as one or two dimensional model of nuclear matter, instead of coming from the realistic reactions. Therefore, our understanding about the real nuclear reaction dynamical processes is not enough and we are still scarcity of the systematic knowledge of the nonlinear characteristics on them. Frankly speaking, we have recognized only the existence of the nonlinearity but not fully grasped the details of chaos in the realistic heavy ion collisions. Therefore, it is particularly important to study such problems nowadays when we do not have a clear definition of integrability or chaos for quantum dynamic systems[17].

 Recently, the multifragmentation of a system in the action of the mean field alone has been observed and the sensitivity of the mean field dynamics to the mechanical instability of the initial distribution of nucleons has been demonstrated by observing the density fluctuation and the Largest Lyapunov Exponent within the Quantum Molecular Dynamical Model(QMD)[18].
In the present paper, taking $^{\prime}$ the essence of chaos is the randomness generated by deterministic nonlinear dynamics$^{\prime}$ as our basic guideline, we are going to re-examine the real dynamical process of the HIC at intermediate energies and observe systematically the manifestations of deterministic chaos in a given reaction system. The Isospin dependent Quantum Molecular Dynamical Model(IQMD) is used to simulate the central collision $^{40}Ca+^{40}Ca$ since the success of the IQMD model in studying the multifragmentation of HICs and the simplicity of this reaction system to be calculated. Of course, the simulation carried out here is easy to apply to other more complicated colliding systems including large systems with peculiar structure or other degree of freedom. The typical characteristic of the chaotic behavior, such as the production of the Generalized Entropy(GE), the fractal structure of the semiclassical phase space are evaluated with the help of the lattice method. The Multifragmentation Entropy(ME) and information dimension, non statistical fluctuation of the fragment mass distribution and the factorial moments in the final state of the reaction are analyzed based on the coalescence model. Our results not only show clearly the specific appearances of such characteristic quantities, but also verify further the existence of chaotic mechanism in the reaction dynamics with the mutually confirmation from several aspects. In the following sections, we will introduce the model briefly and present the features of such quantities in detail.

{\section {Brief review of the dynamical model}}

The IQMD model is an extended version of the QMD[1,2] model in which the correlations have been kept and thus it is very suitable for studying the multifragmentation of HIC at medium energies. Great successes have been achieved in these investigations  by utilizing this model, in particular, in the reproduction of the multifragmentation and explanation of the relevant features of the HICs at intermediate energies induced by radioactive beams [19-22]. Here we recall briefly the main components of the model.

The model contains mainly two ingredients: density-dependent mean field containing correct isospin term including symmetry potential and the in-medium nucleon-nucleon cross sections which are different for neutron-neutron (proton-proton) and neutron-proton collisions.  The potential is
$$
U(\rho)=U^{Sky}+U^{C}+U^{sym}+U^{Yuk}+U^{MDI}+U^{Pauli}, \eqno (1)
$$
where $U^{Sky}$,$U^{C}$,$U^{sym}$,$U^{Yuk}$,$U^{MDI}$ and $U^{Pauli}$ are Skyrme, Coulomb, symetry, Yukawa potential, momentum dependent interaction  and the Pauli potential, respectively. Their concrete expressions and the parameters involved in the formulas are given in Refs.[20-24].

There are a variety of expressions for the nucleon-nucleon(NN) cross section in the model for studying the intermediate energy nuclear collisions. Ref.[25] proposed a simple formula of the isospin dependent NN cross section which includes the medium effect involving the nuclear matter density and incident energy dependencies. It looks like
$$\sigma_{nn}=(13.73-15.08\beta^{-1}+8.76\beta^{-2}+68.67\beta^{4})\frac{1.0+7.772E^{0.06}_{lab}\rho^{1.48}}
{1.0+18.01\rho^{1.46}} \eqno (2) $$
$$\sigma_{np}=(-70.67-18.18\beta^{-1}+25.26\beta^{-2}+113.85\beta)
\frac{1.0+20.88E^{0.04}_{lab}\rho^{2.02}} {1.0+25.86\rho^{1.90}}
 \eqno (3) $$
$$\beta=\sqrt{1.0-\frac{1.0}{\gamma^{2}}}, \gamma=\frac{ E_{lab} }
{931.5}+1.0,        \eqno (4)
$$
where $\beta$ is the ratio of projectile velocity to light velocity and  $\rho$ is nuclear matter density in the unit of $fm^{3}$.  $\sigma_{nn}$ and $\sigma_{np}$ are the neutron-neutron (or  the proton-proton) and the neutron-proton cross sections, respectively. $E_{lab}$ is the incident energy in laboratory frame.

Combining the potential and the NN cross section given in Eqs.(1) to (4), we[26,27] have calculated the isotropy ratio between transverse and longitudinal energies of the free protons moving forward in the center of mass system emitted in the reaction $^{129}Xe$+$^{129}Sn$ in Fermi domain with the IQMD model. And a novel Pauli-blocking factor
$$
\xi(E)=0.644+0.011E-1.513E^{2}+6.214E^{3}   \eqno (5)
$$
has been extracted by comparing the calculations with the experimental data[28]. The factor $\xi$ expressed in Eq.(5) modifies the uncertainty relation $R_{r} \times R_{p} \geq \xi h$  which represents the quantum property of the nucleon in the phase space. Where $R_{r}$ and $R_{p}$ are the radius of the Fermi sphere occupied by a nucleon in coordinate and momentum space respectively. With the combination of the formulas given above the dissipation phenomenon in the heavy-ion reaction at incident energy near Fermi energy have also been studied and some experimental data have been reproduced. In the following investigation, the calculations will be performed with this model including the relevant parameters unchanged.

{\section {Typical characteristics of the nonlinear dynamics}}

{\subsection {Production of entropy and fractal dimensionality }}

Entropy is an important thermodynamic quantity. The variation of its' magnitude reflects the confusion in the microscopic state of a system, or the equilibrium of the system under certain macroscopic condition. And thus the entropy production in the HICs has attracted considerable attention in both nuclear physics and nonlinear dynamics [12,29]. In 1981, Bertsch and Cugnon[30] studied quantitatively the entropy production in the collision of $^{40}Ca+^{40}Ca$ at incident energy $E=800MeV$ within cascade model and concluded that the generation of the entropy is closely related to the formation of clusters. In Ref.[31], based on the fireball model, the authors investigated the same issue and speculated that the amount of the entropy  production is not the same during the various stage of nuclear reactions and pointed out that there is barely entropy generation in the final stage, expansion stage, since the density of the particle has become to be so small, thus they seldom collide and the Liouville's theorem guarantees that the density of particles in phase space remains constant in the absence of collisions. However, S. Das Gupta et. al. [32] developed a microscopic model for treating the fragmentation of nuclear matter in which both hard collision and propagation of nucleons are treated. They argued that the mean field can lead to fragmentation of nuclear matter and increase of the entropy simultaneously. Therefore, the mechanism responsible for, and the significance of, the entropy generation in HICs have been a matter of much disputed issue in nuclear physics[29]. Here we want to observe the variation of the entropy in the real nuclear reaction instead of the schematic model.

The entropy for a Fermi system is defined by[31]
$$
  S=-\int d\gamma [{ f\ln( f)-(1-f)\ln (1-f)}],~~~~~~~~   d\gamma=g \frac{d{\vec r}d{\vec p}}{(2\pi\hbar)^{3}}           \eqno (6)
$$
where $f=f({\vec r},{\vec p})$ stands for the occupation probability of nucleon in a phase space volume of $h^{3}$ and $g=4$ is the degeneration of the spin-isospin degree of freedom.  In order to avoid the complicated integration which is very time-consuming, we turn to calculate the time evolution of the coarse-grained entropy, namely, the so-called generalized entropy (GE) defined in Ref.[32] and adopt the ansantz used in Ref.[33]. Specifically, the available phase space is broken up into cells of volume $\beta h^{3}$ with $\beta$ being an adjustable parameter. Then the entropy becomes
$$
S= -4 \beta \sum_{i} [n_{i}\ln(n_{i})-(1-n_{i})\ln{(1-n_{i})}],~~~~~~n_{i}=\frac{N}{4R\beta}          \eqno (7)
$$
where $N$ is the number of nucleons in a cell of volume $\beta h^{3}$ and $R$ the number of runs in our numerical simulation. $n_{i}$ indicates the occupation probability of a phase space of volume $h^{3}$.
However, it is still not easy to perform the calculation of Eq.(7) in the six dimension space, in particular for a nucleus with large numbers of nucleons. Moreover, there is an uncertainty in the amount of $S$ which stems from the adjustability of the parameters, $\beta$, $R$ and the size of cells. Fortunately, Ref.[32] has proved that the trend of the variation of GE does not depend sensitively on the size of cells although the magnitude of entropy is determined significantly by the volume of a cell. Coincidentally, our main concerning focus mainly on the tendency of GE variation reflecting the chaotic behavior of the system. Therefore, we calculate the generation of GE in the central collision of $^{40}Ca+^{40}Ca$ at incident energy 800 MeV/u and the results are depicted in the Fig.1. It shows that the value of the entropy is saturated at $4.0-4.4$, which is a reasonable value quoted in Ref.[30] for 10 events or 800 nucleons. In the left panel of the Fig.1 the curves from bottom to top correspond respectively to the magnitude of mesh side-length in six dimension space, (a) $\Delta r=35 fm$ , $\Delta p=0.29 GeV/c $, (b) $\Delta r=24 fm$ , $\Delta p=0.19 GeV/c $, (c) $\Delta r=17 fm$, $\Delta p=0.15 GeV/c $ and (d) $\Delta r=14 fm$, $\Delta p=0.12 GeV/c $  with $\beta=0.8$  in each cases.
Here the $\Delta r=max \{\Delta x,\Delta y, \Delta z\} $ and $\Delta p=max\{\Delta p_{x},\Delta p_{y}, \Delta p_{z}\} $ are the largest side length of the mesh in coordinate and momentum, respectively. In the actual calculations, $\Delta x \neq \Delta y\neq  \Delta z$ since the spread velocity of nucleons is not the same in different direction at a given time. The division of the momentum space is alike in coordinates space.
 Comparing the behavior of each curve we can learn that the sensitivity of the variation of the GE to the mesh size since the function $n_{i}\ln( n_{i})$ depends sensitively on the value of the distribution $n_{i}$, as has illustrated in the Fig.1 of Ref.[30]. The left panel in our Fig.1 shows that the magnitude of the entropy tends to saturation with decrease of the size of the cell. Of course, It can be expected that the location and the magnitude of the peak of $S$ will vary  with the size of cells if we keep the size going down so that the most cells would be empty. But we are not going to do so because we have come to a reasonable range in the case (d) with the $S \approx 4.0 \sim 4.4 $ which is consistent well with the value given in Ref.[30]. We also notice that the increment $\Delta S \approx 0.1 \sim 0.2$ produced in the expansion stage of the reaction[33] has been confirmed in our calculation.
\begin{center}

  {\bf  Fig.1}

\end{center}
Our results, obtained by using IQMD and the lattice method, demonstrate dynamically the production of the GE during the whole reaction process, i.e., the rapid increase in the stage of the collision phase and the slow growth in the expansion stage of the reaction. We can easily exhibit further the generation of GE in the coordinate space and the momentum space, separately, shown in the right panel of the Fig.1 obtained with any model parameter involved unchanged in the previous calculations. It is clearly seen that the GE grows much substantially in the momentum space (dash curve) than that in the position space (solid curve) in the compressed stage of the reaction where the bombarding energy converts rapidly into the thermal energy. In order to visualize the thermalization of the reaction process, we explore the averaged transverse momentum transfer(TMT) of the colliding system as has done in Ref.[34]. To this end it is needed first to determine the reaction plane by defining a vector
$$ {\vec Q}=\sum^{M}_{\mu} \omega {\vec p}^{t}_{\mu}              \eqno (8) $$
where ${\mu}$ is a particle index and $ \omega_{\mu}=+1 $ for nucleons with longitudinal momentum $ p_{z} \geq 0$ and $ \omega_{\mu}=-1 $ for $ p_{z} < 0 $. $M$ is the number of the nucleons with rapidity in the range of $-1.5 \leq Y \equiv \frac{1}{2} \ln ( \frac{ E+p_{z} }{ E-p_{z} } ) \leq 1.5 $.  The momentum of nucleon $\nu $ projected on the reaction plane, $ p^{X}_{\nu} = { \vec p }^{t}_{\nu} \frac{ {\vec Q }_{\nu} }{| {\vec Q}_{\nu}|}$,  can be used to estimated the average transverse momentum transfer in the reaction. Where ${\vec Q}_{\nu} ={\vec Q}-{\vec p}^{t}_{\nu}  $ and ${\vec p}^{t}_{\nu}=p_{x}(\nu){\vec i}+p_{y}(\nu){\vec j}$. The variation of averaged values $<P^{X}>$  for per particle in each event with respect to the time is plotted in Fig.2. It shows that the average transverse momentum transfer increases rapidly up to its summit at $t^{TMT} \simeq 16 (fm/c)$ and then become nearly an constant. This indicate the relaxation time reflected by transverse momentum transfer is approximately consistent with that of the GE rapidly increase, $t^{GE}_{rap} \approx \ 22 fm/c$, in the momentum space showing by Fig.1,comparing to the whole reaction time. The subtle differences between $t^{TMT}$ and $t^{GE}_{rap}$  can be explained by the statistical error in the calculations for their averaged values according to the different mathematical expressions. This approximate coincidence confirms further the conclusion drawn in Ref.[34] that the transverse momentum transfer should be viewed as a convenient tool which is complementary to entropy analysis and allow us to estimate the degree of thermalization of the reaction process.

\begin{center}

  {\bf  Fig.2}

\end{center}

Compared the rate of GE productions in different spaces, we can see that the transverse momentum transfer of the nucleons is much rapid than the diffusion of the nucleons in spatial space. And thus the change of GE in momentum space is faster than that in the position space since the increase rate of the GE is proportional to the rate of particle spread out. To be more precise, the increase of GE in the collision stage  mainly comes from the thermalization of the colliding system, while the generation of GE in the expansion stage of the reaction stems mainly from the diffusion of the composite system.

It is worth to notice that the GE decrease slightly in the momentum space due to the reduce of the density in the momentum space with time, while it increases continuously in position space with respect to time in the expansion stage of the reaction. Nevertheless, the total amount of the GE is still increase slowly in the phase space showing by the curve (d) in the left panel of Fig.1. This behavior indicates that the expanding of the region in position space is partly compensated by the shrinking of the velocity distribution into a smaller region in the momentum space as the matter cools.

As has pointed out by Schuster[35] that the entropy increases not only due to the increase of the particle number, but also due to the dynamical fluctuations. We consider a multifractal description of that fluctuation and focus on the information dimension as a characterization of the entropy of the system. Fractal geometry is also a new science which devotes to answer the question how the microscopic behavior is related to what we observe on the macroscopic scale. According to the Fractal theory the information dimension $D_{I}$ is defined in[36]
 $$
 D_{I} = - \lim_{\delta \to 0} \frac{S}{ ln(\delta)}           \eqno(9)
 $$
with $\delta$ being the size of each cell.

In the Fig.3 we show the result for the central collision of
$^{40}Ca + ^{40}Ca$ at incident energy $E=800 MeV/u$ at time $t=100
fm/c$ when the generalized entropy has come to a plateau. In this
figure the straight line is the fitted results with our computed
results indicated by the scattering points and the extracted
information dimension is $D_{I}=0.484$. $D_{I}$ is not an integer
meaning the existence of the self-similarity in the distribution of
the available phase space.

\begin{center}

  {\bf  Fig.3}

\end{center}

In 1999, Y.G. Ma introduced a method [37] to diagnose a nuclear liquid gas phase transition by multiplicity entropy(ME),
$$
H=k_{b} \sum_{i} p_{i} \ln (p_{i}),                  \eqno(10)
$$
which determines the critical point by finding the maximum value of multiplicity entropy in a certain state of the system. In this definition $k_{b}$ is Boltzmann constant and the probability distribution $p_{i}$ is the ratio of the number of `i` particle produced by $ N_{i} $  to the total number of particles produced by $N$, i.e. $p_{i} = N_{i} / N $ and $\sum_{i=1}^{N}p_{i}=1$. He used the ME to study the phase transition of nuclei in the framework of the isospin dependent lattice gas model and molecular dynamical model, and concluded for the first time that the maximum of the ME indicates that the system comes at a largest fluctuation/stochasticity/chaoticity in the event space. It is naturally making sense to check such behavior of the ME and compare it with the GE in the given reaction system under exactly the same incident conditions. Our calculated results for the ME for the reaction system considered here is presented  in the left panel of Fig.4.
\begin{center}

  {\bf  Fig.4}

\end{center}

This figure shows that the maximum of the ME
appears at the time $t^{ME}_{Max} \approx 16 fm/c $, and the
multiplicity reaches its' maximum value at about $ t^{Nimf}_{Max} \approx 22fm/c $ showing in
the right panel of the Fig.4. The discrepancy between $ t^{ME}_{Max} $ and $ t^{Nimf}_{Max}$  is mainly caused by their different mathematical expressions
since they are obtained by the analysis of the same set of simulation results.
Therefore, the appearance of ME maximum reflects the formation of primary fragments and a phase transition from two colliding nuclei to multiple clusters,
i. e. the  multifragmentation phase transition. In the terminology of nonlinear dynamics, it is
a transition from order to disorder of the nucleon motion involved in the reaction.

Comparing the behavior of the variations of the GE shown in the
Fig.1 and the ME in Fig.4, we can see that both kinds of
entropies have a common feature that they increase rapidly in the
compress stage of the reaction. But there is a slight difference in time when the two kinds
of entropies reach their respective maximum. This can be understood with the help of the behavior of
the transverse momentum transfer which increase rapidly with time in this reaction stage, shown in the Fig.2.
Meanwhile, in our calculation the clusters are constructed by
using the coalescence model[38],  which requires that the distance
in phase space between nucleons belonging to one cluster  must be
less than $P_{0}$ and $R_{0}$ simultaneously. Specifically,
$P_{0}=350 MeV/c $ and $R_{0}=0.3 fm$ here. In
other words, the distance between nucleons belonging to different
clusters must be greater than P0 and R0. However, in the calculations for the generalized entropy
performed before the size of the lattice cells, e. g. $\Delta r=14 fm$, $\Delta p=120 MeV/c $ for the curve (d) in Fig.1,  is not exactly equal to the values of $P_{0}$ and $R_{0}$.
Namely, the nucleons located in one lattice cell may belong to different clusters and vice versa. Therefore, the time when the maximum of the multifragmentation entropy reached is not exactly same as that when the generalized entropy reaches its' maximum. In our numerical results $ t^{ME}_{Max} < t^{GE}_{rap} $.
Anyway, the production of both types of the entropy is
responsible for the increase in the disorder or the chaoticity of nucleons motion.

{\subsection {Factorial moment and intermittency} }

 Intermittency is a manifestation of the scale invariance of a physical system or a nonlinear dynamical process and the randomness of underlying scaling law. According to nonlinear dynamics theory [39], the emergence of the intermittency during the spatial-temporal evolution of a dynamics system is considered as a indicative sign of chaotic behavior, similar to the period-doubling bifurcation. The original idea of the studying intermittency in nuclear collisions came from the work of Bialas and Peschanski [40] who looked at the rapidity distribution of produced particles in cosmic ray experiments. They proposed using scaled factorial moments of the rapidity distribution to study the possible appearance of intermittent behavior in such collisions and evidence for non-Poissonian fluctuations. The greatest merit of this method is that the factorial moments filter out the statistical fluctuation and retain only dynamical fluctuation which is just the most essential concern of the chaotic dynamics. The intermittent behaviour and the self-similarity pattern of the fluctuations in the charge distribution in the breakup of $^{197}Au + ^{179}Au$ at energy $E=1GeV/u$ in a nuclear emulsion[41] have been explained by scaled factorial moment method based on the percolation model[40,41,42] and by the microcanonical model[42,43]. For the purpose of finding the critical phase transition signal, references [42,44,45] have studied the normal and factorial moments of fragment distribution in heavy ion collisions near the Fermi energy by quantum and classical dynamics  models respectively and pointed out that the intermittency signal is most obvious near the critical phase transition point. As a typical signal of nonlinear dynamics and self-similarity of complex systems, here we quantitatively investigate the general performance of the factorial moments in our nuclear collisions.

 The factorial moment defined by Bialas and Peschanski[40] is
$$  F_{q}=M^{q-1} \frac{\sum^{M}_{m=1}n_{m}(n_{m}-1)...(n_{m}-q+1)} {N(N-1)...(N-q+1)}               \eqno (11) $$
Where the range $\Delta A$ of the distribution of fragment mass $A$
is divided into $M$ bins with each of interval $\delta
A=\frac{\Delta A}{M}$. $N=\sum^{M}_{m=1}n_{m}$ is the total number
of fragments with $n_{m}$ being the number of the fragments in bin
$m$. The factorial moment for the central collision of $^{40}Ca +
^{40}Ca$ at incident energies $E=800MeV$ is plotted in Fig.5. Where
the results are obtained in the simulation of our colliding system for 1000 events.

\begin{center}

  {\bf  Fig.5}

\end{center}

Each curve corresponding to different $q$ in the Fig.5 looks roughly like a straight line indicating a power-like increase of the factorial moment with decreasing bins length. The slops of these fitted lines, $\alpha_{q}$, termed as the intermittency exponent[40,43], can be express as
$$F_{q} \propto (\delta Z)^{-\alpha_{q}}.    \eqno (12) $$
This is just the typical characteristics of the self-similarity[46] emerged at all scales in the distribution.  The fractal dimension can be derived from the formula
$$d_{q}=\alpha_{q} / (q-1)).    \eqno (13) $$
Both the intermittency exponent $\alpha_{q}$ and fractal dimension $ d_{q} $ for various $q$ are listed in Table I.
\begin{center}
 Table 1: the intermittency exponent and fractal dimension
\begin{tabular}{|l|c|c|c|c|r|}
\hline
q&2&3&4&5&6\\
\hline
$\alpha_{q}$~~~ &0.524&1.138&1.765&2.393&3.022\\
\hline
$d_{q}$ &0.524&0.569&0.588&0.598&0.604\\
\hline
\end{tabular}
\end{center}
As showing by the table I the fractal dimension $d_{q}$ is not integral indicating that there still is fractal structure in the fragmentation
 pattern though the incident energy is much higher than Fermi energy and the collision between nucleons more violence comparing to that considered
 Ref.[44-47]. Besides, there are a few characteristics worthy of our attention:
 (1) The factorial moment exponents increase with their order 'q' indicating that the phase space structure of multifragmentation is a multifractal structure;
 (2) it is impossible to infer that whether the fragmentation process of the colliding system is a prompt fragmentation or sequent decay[42] due to the dimensions
 being neither constant nor directly proportional to $q$.

We have actually completed the calculations and analyses of the
factorial moments for the colliding system considered here at
different incident energies varying from $E=50MeV$ to $E=800MeV$.
The results are similar to that showing in the Fig.5 except the
magnitudes of the intermittency exponents increasing with the
incident energy. Since the second factorial moment relate to the the
mean value $<n_{m}>$ and the variance $\sigma_{m}$ of the
multiplicity distribution in the $m$th bin, explicitly, $F_{2}=1+
\frac{ \sum_{m}(\sigma^{2}_{m}-<n_{m}>) } { \sum_{m} <n_{m}>^{2} }
$, and hence we only plot the variation of the slop of $F_{2}$, i.e.,$\alpha_{2}$,
as a function of the incident energy in Fig. 6.
\begin{center}

  {\bf  Fig.6}

\end{center}
This figure shows that the intermittency exponents increases monotonically with the incident
energy though the rate of the increase varying with the energy.

{\subsection {Power Law in the fragments mass  distribution }}

It is generally believed that nuclear fragmentation occurs near the critical point of liquid-gas phase transition of nuclear matter based  intensively on
the studies both experimentally and theoretically. According to Fishers[48,49,50] semi-phenomenological droplet model for equilibrium liquid-gas transitions
in infinite systems, the fragment mass distribution should be a power law multiplied by an exponential whose argument goes to zero at the critical point:
$$ Y(A)=q_{0} A^{-\tau} \exp{ \{A\Delta \mu - c_{0} \epsilon A^{\sigma}/ T  \}}   \eqno(14) $$
where A is the mass of the drop, $\Delta \mu$  is the difference of the chemical potentials of the gas and liquid phases, $\sigma$  is the surface tension. At the critical point $\Delta \mu$, $c_{0}$  and $\sigma$  go to zero and we are left with a simple power law,
$$ Y(A)=q_{0} A^{-\tau}           \eqno(15)$$
For infinite systems in equilibrium the value of the exponent $\tau$ lies in $2<\tau<3$.

Ones have found that the power-law distribution exists universally in natural and social phenomena such as the distribution of citation of scientific papers[51].
 At present there is no definite answer to the question about the dynamical mechanism of power-law distribution.
 The critical phase transition is one of the various speculated mechanisms. The existing studies have pointed out that the power-law distribution is a necessary
 condition for the critical phase transition, but is not a sufficient condition for a system being in equilibrium. In particular,
 the power-law behaviour can be observed in the non-equilibrium phase transition in the self-organized critical phase transitions which is a process of entropy
 reduction and ordering. In the context of nonlinear complex systems, the power-law distribution can be used as the evidence of the critical state of
 self-organizing systems[52], that is, a sign of the transition from steady state to chaotic state.
 Therefore, it is equally important for us to understand the dynamics mechanism of heavy ion collisions in nuclear physics and the generation mechanism
 of power-law distribution in nonlinear dynamics.

Here, again, we simulate the colliding system, $^{40}Ca + ^{40}Ca$,
due to its simplicity
to explore the general features of the distribution of the intermediate mass fragments(IMF) in the final state of this reaction.
The distributions of the fragments with masses in the range of $ 2 < A \le 10$ have been analyzed for the incident energies varying from $E=50 MeV$ to
$E=800 MeV$ and the power law behaviors in these distributions have been checked. In
 Fig.7 we give the IMFs distribution( isolated square points ) and the fitted results(straight lines) at $E=100 MeV$(up panel) and $E=800 MeV$
(down panel). The resulting power law exponents are $\tau = 2.17$ for $E=100 MeV$ and $\tau = 5.02$ for $E=800 MeV$, respectively.
In order to test the reliability of our fittings, we sample a set of
100000 data according to the standard power-law distribution
function, $Y(x) \sim  x^{-\tau}$ with $\tau$ taken as the corresponding fitted values, and compare the complement cumulative
distribution functions(CCDF) of these standard models with that of our fitted
data, so as to obtain the Kolmogorov Smirnov(KS) statistic[53],
which is simply the maximum relative distance between the CCDFs of
the models and the fitted data,
$$D=Sup [\frac{ |S(x)-P(x)| }{ \sqrt{ P(x)*(1.0-P(x)) } } ].      \eqno (16) $$
Here $S(x)$  and  $P(x)$ are the CCDFs of the data and the
power-law model, respectively. The comparison is, as an example, presented in Fig.8 for $E=100 MeV$  and
the estimated $D=0.14$.

\begin{center}

  {\bf  Fig.7}

\end{center}

\begin{center}

  {\bf  Fig.8}

\end{center}
The variation of the power-law index(left pannel)
and the KS statistic(right pannel) for the colliding system at
various incident energies from 50MeV to 800MeV are depicted in Fig.9. It shows that the
fitted power-law index and the corresponding KS statistics go up
with the increase of the bombarding energy and the latter indicates the distribution come from our numerical fittings become more
 inconsistent with the their corresponding standard power-law distributions in the course of energy increasing.
It should be especially  pointed out here that our conclusion on the power-law distribution is obtained under the premise of
a given collision system(40 nucleons), fixed fragment mass interval($ 2 < A \le 10$) and the number of simulation events(1000 runs) in the whole energy range considered to
obverse the general characteristics of the nonlinear dynamics. Therefore, our calculation results are not enough to exact the most appropriate power law exponents,
i.e., the minimum index. Such study is currently under way.

\begin{center}

  {\bf  Fig.9}

\end{center}

In order to exmine intuitively the difference of the multifragmentation of the colliding systems
at various incident energy, we plot the largest fragments at different incident energies and analyze
the normalized variance of the masses of the largest fragments(NVM), which is an alternative signal for testing the gas-liquid phase transition,
proposed in Ref.[54]. The definition of NVM is
$$ \sigma_{NVM}=\frac{<A_{MAX}^{2}>-<A_{MAX}>^{2}}{<A_{MAX}>}                \eqno (17)  $$
where $A_{MAX}$ is the mass of the largest fragment and the symbol $<.>$ stand for the average over an ensemble of events at a given energy.
The study of Ref.[54] have pointed out that there will be a maximum of the NVM in the critical region
of a phase transition system. For our reaction, the changes of the NVM with increase of the incident energy are plotted in the left panel of Fig.10 and the maximum of the NVM emerges at $E=100 MeV$ perspicuously.
The average number of the largest fragments continues to decline with the increase of incident energy, showing in the right panel of Fig.10. Taking the appearances of this figure and Fig.9 together, we can learn that
the characteristics of critical phase transition are the most obvious near $E=100 MeV$ in the collision system and the energy region considered in this paper.

\begin{center}

  {\bf  Fig.10}

\end{center}

{ \section {Summary and discussion} }

In the present paper, we have investigated systematically the
chaotic characteristics of the multfragmentation in the heavy ion
collisions at intermediate energies by simulating a  relatively
simple and representative collision system. Since entropy is a
measure of the degree of chaos of the system, the generation of
entropy marks the transition of the system from order (low symmetry)
to disorder (high symmetry). we have studies and compared the
productions of generalized entropy in the reaction process and the
fragmentation entropy of fragments distributions in the final state
of the reaction. The former reflects the characteristics of nucleons
distribution in semiclassical phase space in the collision process
and the latter indicates the feature of the final products of the
reaction in the event space.  Our results show that the disorder of
the system increases rapidly in the compression stage of the collision. The production of the generalized entropy in the momentum space
corresponds to the thermalization of the system in such process the direct motion of the nucleons
converts to random thermal motion. The time when the generalized entropy reaches a plateau in
the course of evolution in the momentum space is nearly the same
as that when the maximum of the multifragmentation entropy
reaches. These coincidences demonstrate that the production of the
entropy mainly occurs in the process of fragment formation. The
influence of the mean field on the generalized entropy increases
slowly while the multifragmentation entropy decreases gradually with
respect to the separation of fragments.

The critical theory  has demonstrated that the correlation length
between the clusters or produced particles tends to be infinity near
the critical point is the dynamic origin of scale invariance leading
to the fractal. Numerically, the appearance of the fractal is the
dimensionality of the trajectory of a nonlinear dynamic system in
space is fraction rather than integer, which reflects the
self-similarity and the degree of chaos of a system [55]. Our
results show that the fractal exists not only in the phase space
distribution of nucleons in the process of heavy ion reaction, but
also in the fragments distribution of the final state of the
reaction.

Intermittency is another  manifestation of the scale invariance of
the physical process and the randomness of underlying scaling law.
It is also a path transiting from regular to chaotic motion for a
complex system. Our numerical analyses for the factorial moments and
the corresponding exponents indicate that the structure of the phase
space is a multifractal for the multifragmentation of the system and
the multifragmentation is neither a simple prompt process nor a
simple sequent decay but both of these processes occur
simultaneously.

As stated before,  power law of the fragment distribution in heavy
ion collisions at intermediate energy is the direct consequence of
Fishers theory for the infinite nuclear matter at critical point.
And thus, sometimes, it can be used as a criterion for the
equilibrium critical phase transition of a system. However, the
researches in self-organized critical theory, developed in recent
years, show that the emergence of power-law distribution is neither
the necessary condition for a system to be in equilibrium, nor the
evidence for the system to be at its' phase transition critical
point. For instance, the frequency of occurrence of unique words in
a novel [51]. Moreover, another challenge is to fit a given
distribution, which is only subsets of much larger entities, into
power law and the extraction of the power law exponents
numerically[53]. Here we take the power law as an indicator of the
self-similarity of the nonlinear dynamical process and obverse its
concrete manifestation in the heavy ion collision in a large range
of incident energy. The power-like law has been observed in the
final state of the reaction considered here within the energy range
from near Fermi energy to 800 MeV. Nevertheless, the agreement
between our calculated results and the standard power law model
becomes worse and worse with the increase of the bombarding energy.
Relatively speaking, the characteristics of critical phase transition are the most obvious
in the vicinity of bombarding energy $E=100 MeV$.

In short, ones have studied the heavy ion collisions  and
multifragmentation at intermediate energy for decades and great
achievements have been made. In particular, most of experimental
observations have been explained correctly with appropriate
theoretical model, including the explanation of the scaling
properties of the multifragmentation e.g. Ref.[56]. However, it is
still a contemporary subject to study the dynamic mechanism of the
multifragmentation and the critical behavior of the phase transition
of finite nuclei[57,58], and to search for the physical observable that can
mark the peculiar nonlinear dynamical process since the collision is
a complex quantum many-body process. Taking a relatively simple and
representative realistic collision system as an example, in this
paper, we have studied the typical chaotic characteristics of the
heavy ion collisions at intermediate energies from the perspective
of dynamics and statistics. The results obtained from these two
views confirm each other and explain jointly the chaotic dynamic
mechanism of the multifragmentation phase transition. Accordingly,
we have gotten a clear image of chaotic dynamics of the
multifragmentation in heavy ion collisions since we have clarified
the specific performances of such characteristic quantities.

\vspace{2cm}

\vskip 1.0cm

\section*{Acknowledgements}

This work is supported by the National Nature Science Foundation of
China under Grant No.11665019 and Natural Science Foundation of Gansu Province under Grant No. 20JR5RA497.

\newpage

\begin{figure}[htpb]
\centerline{ \epsfig{file=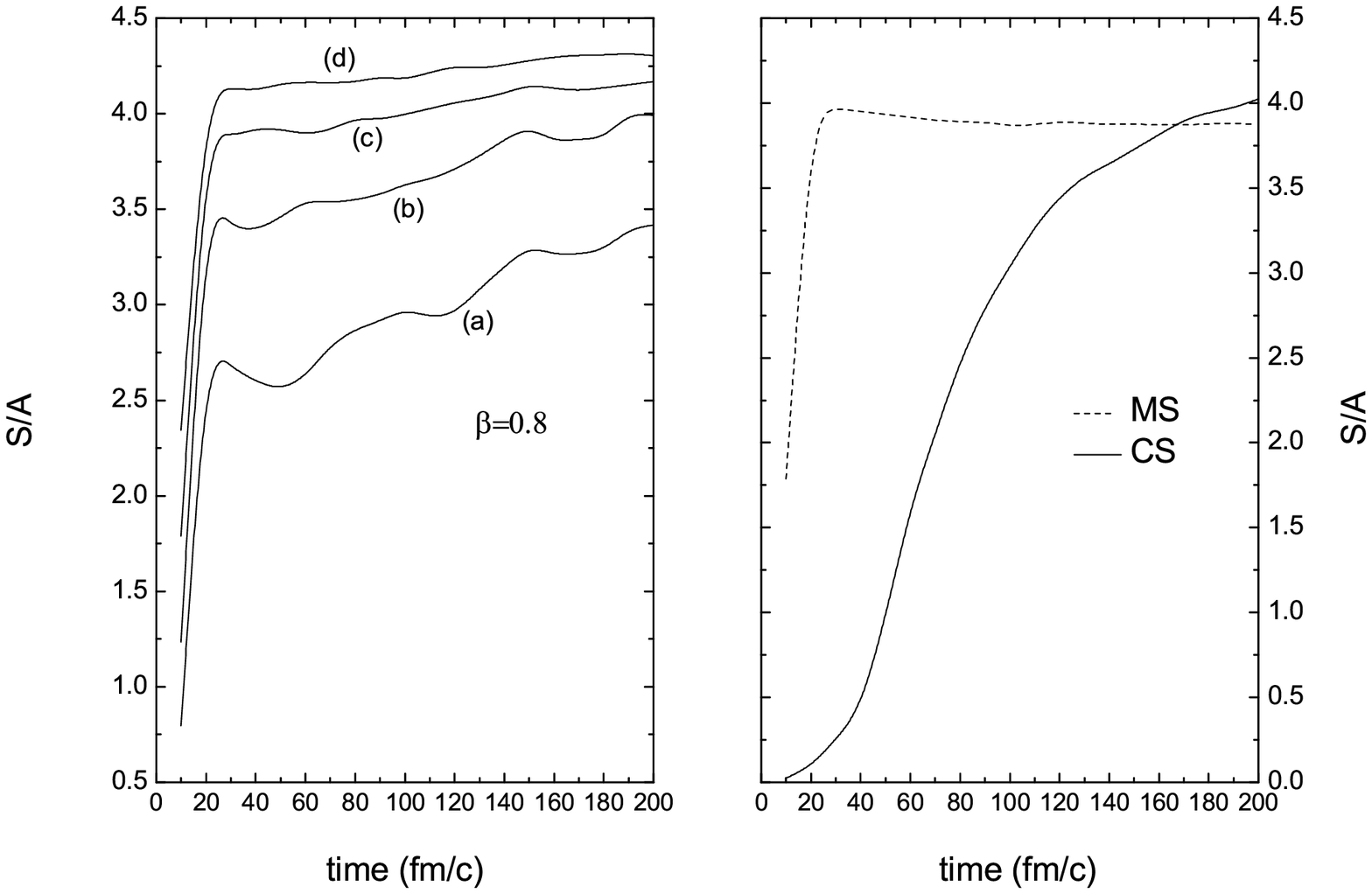,width=20cm,height=12cm,angle=0} }
\caption{The variation of the Generalized Entropy (GE) with respect
to time in the central collision of $^{40}Ca+^{40}Ca$ at incident
energy $E=800 MeV/u$.
 Left panel: the GE created in full phase space. Right panel: the GE created in position space (solid line) and in the momentum space (dashed line). }
 \label{Fig1}
\end{figure}

\begin{figure}[htpb]
\centerline{ \epsfig{file=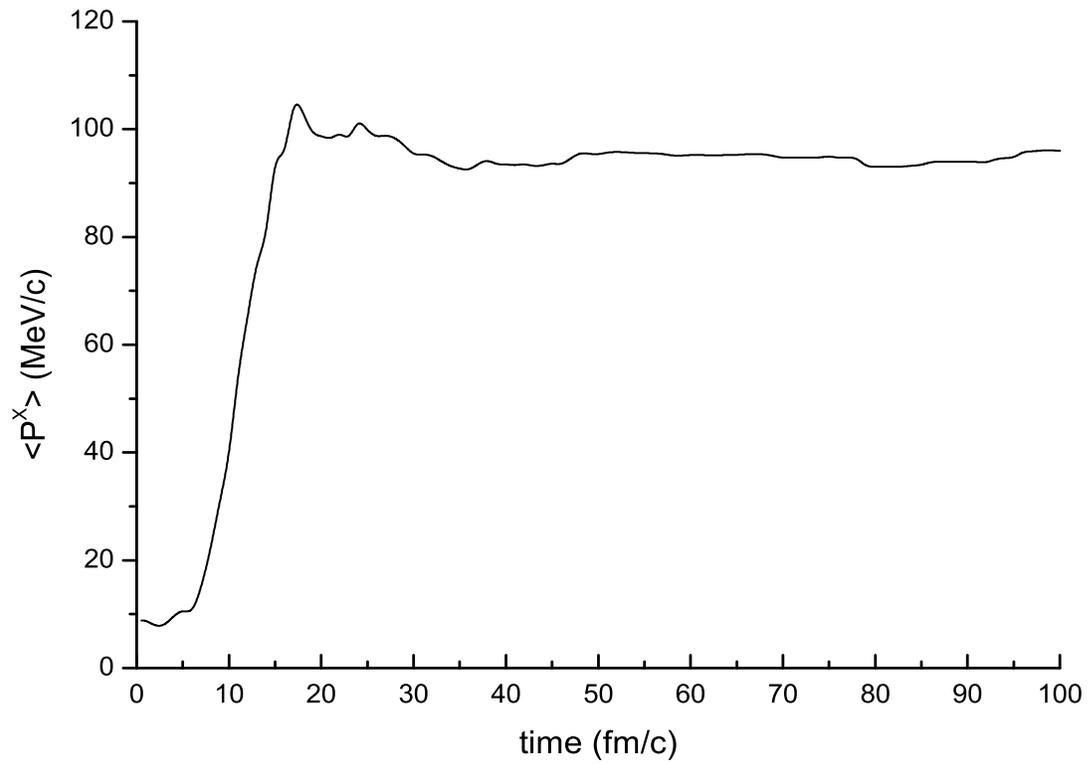,width=18cm,height=12cm,angle=0} }
\caption{The averaged transverse momentum transfer of the colliding
system as a function of time in the same collision as given in Fig.1. }
 \label{Fig2}
\end{figure}

\begin{figure}[htpb]
\centerline{ \epsfig{file=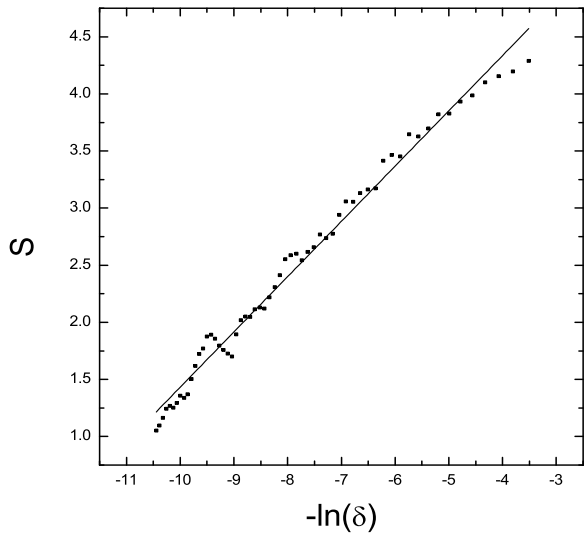,width=18cm,height=12cm,angle=0} }
\caption{The information dimension corresponding the Generalized
Entropy at time $t=100 fm/c $ in the same collision as given in
Fig.1.}
 \label{Fig3}
\end{figure}

\begin{figure}[htpb]
\centerline{ \epsfig{file=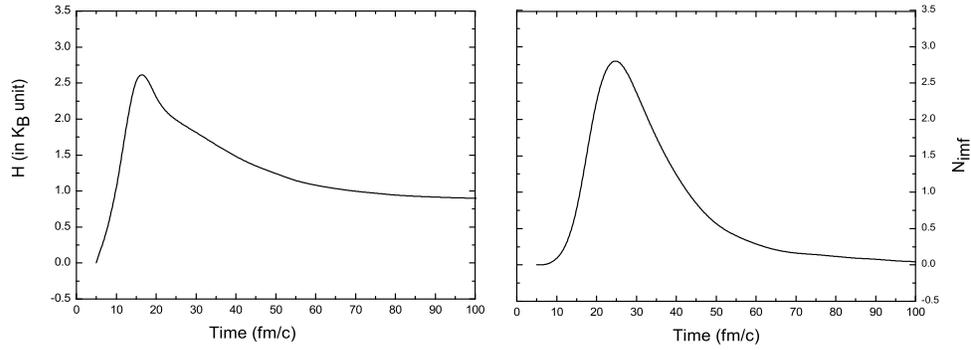,width=18cm,height=12cm,angle=0} }
\caption{The variation of the Multifragmentation Entropy(Lift) and the multiplicity of intermediate mass fragmentation(Right) in
term of the time in the central collision of $^{40}Ca + ^{40}Ca$ at
incident energy $E=800 MeV/u$. }
 \label{Fig4}
\end{figure}

\begin{figure}[htpb]
\centerline{ \epsfig{file=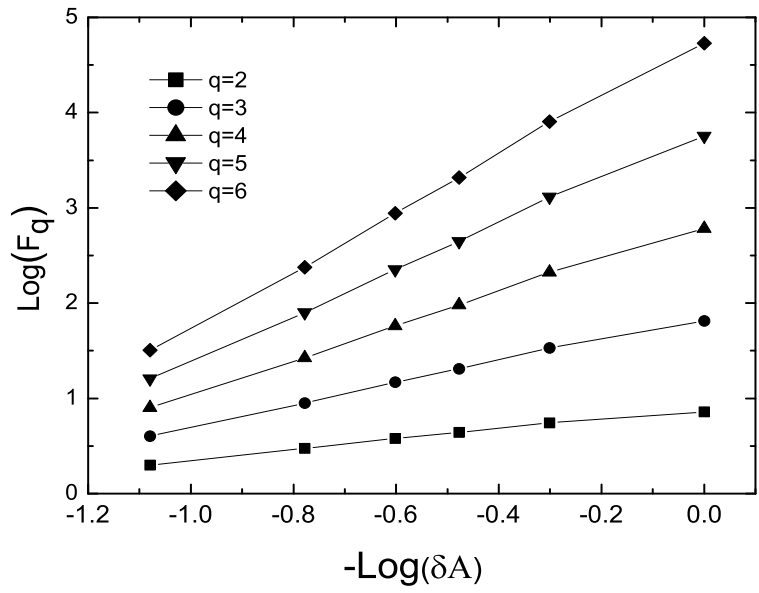,width=18cm,height=12cm,angle=0} }
\caption{The factorial moments for the central collision of $^{40}Ca
+ ^{40}Ca$ at $E = 800 MeV/u$. }
 \label{Fig5}
\end{figure}

\begin{figure}[htpb]
\centerline{ \epsfig{file=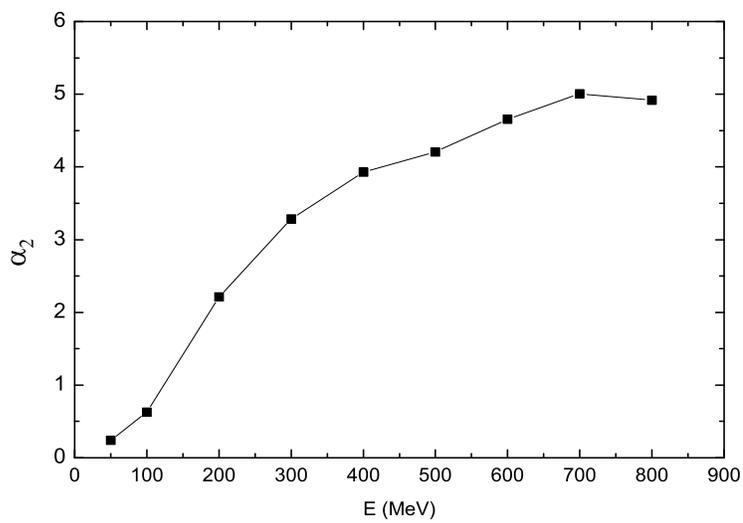,width=18cm,height=12cm,angle=0} }
\caption{The slop of the second factorial moment as a function of
the incident energy.}
 \label{Fig6}
\end{figure}

\begin{figure}[htpb]
\centerline{ \epsfig{file=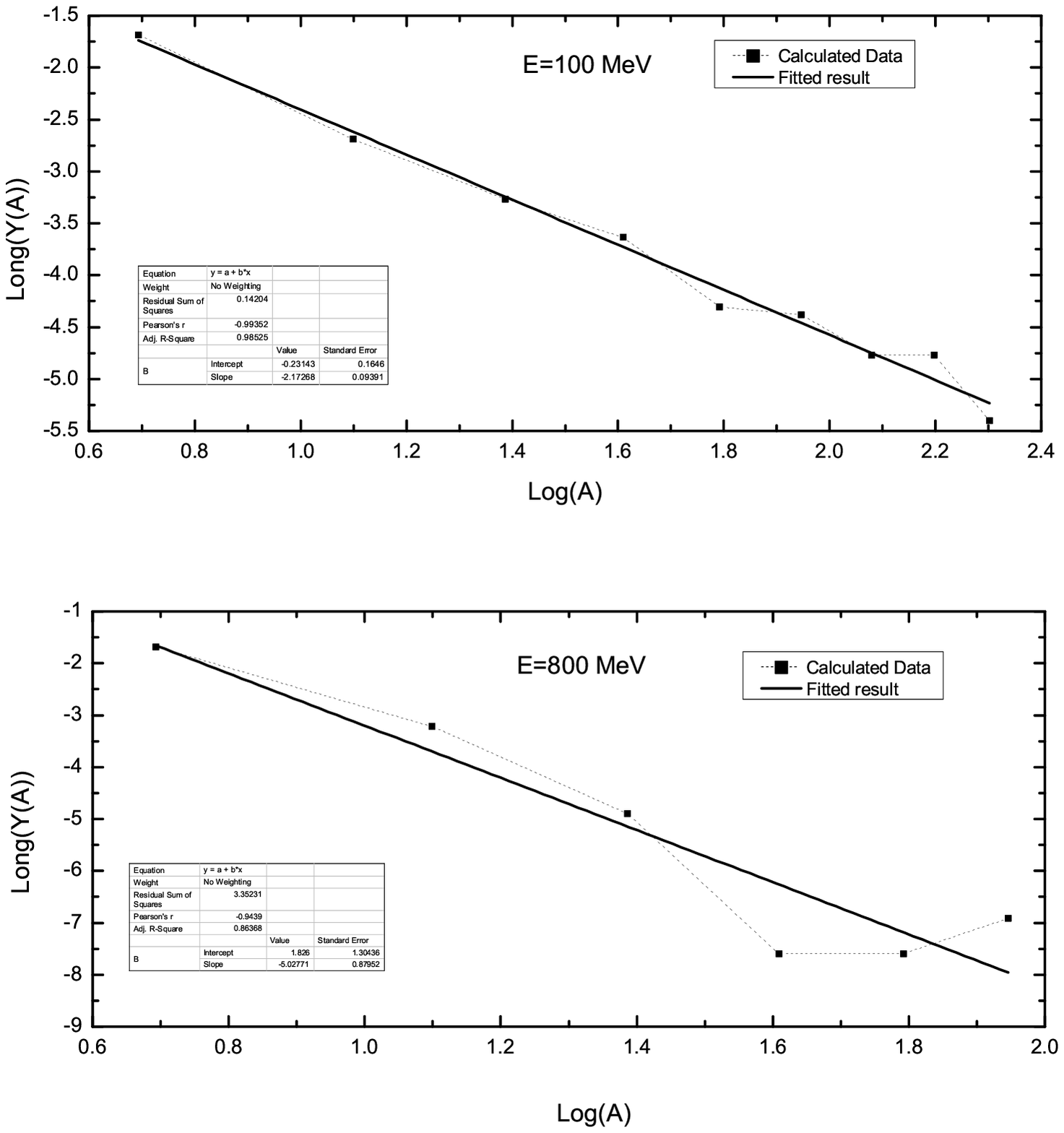,width=18cm,height=12cm,angle=0} }
\caption{The power law distribution of the intermediate mass
fragments, $2 \le A <10$ produced in the central collision of
$^{40}Ca + ^{40}Ca$ at $E = 100 MeV/u$(upper panel) and $E = 800 MeV/u$(lower panel). }
 \label{Fig7}
\end{figure}

\begin{figure}[htpb]
\centerline{ \epsfig{file=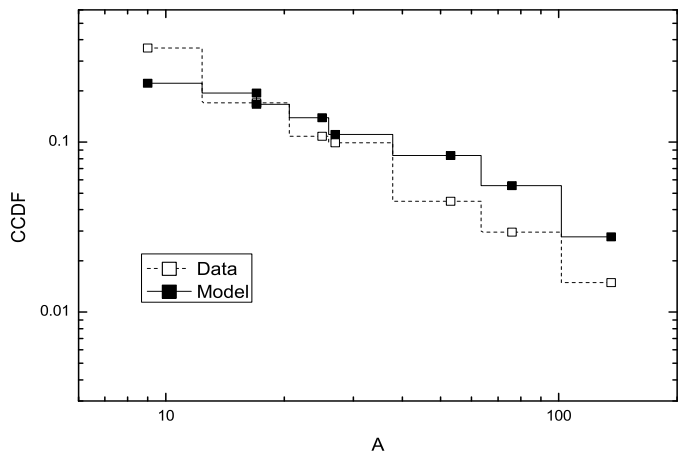,width=18cm,height=12cm,angle=0} }
\caption{The comparison of the CCDFs of the standard power law
distribution( indicated by dashed-line) with that of the data(solid
line) shown in Fig.7. }
 \label{Fig8}
\end{figure}

\begin{figure}[htpb]
\centerline{ \epsfig{file=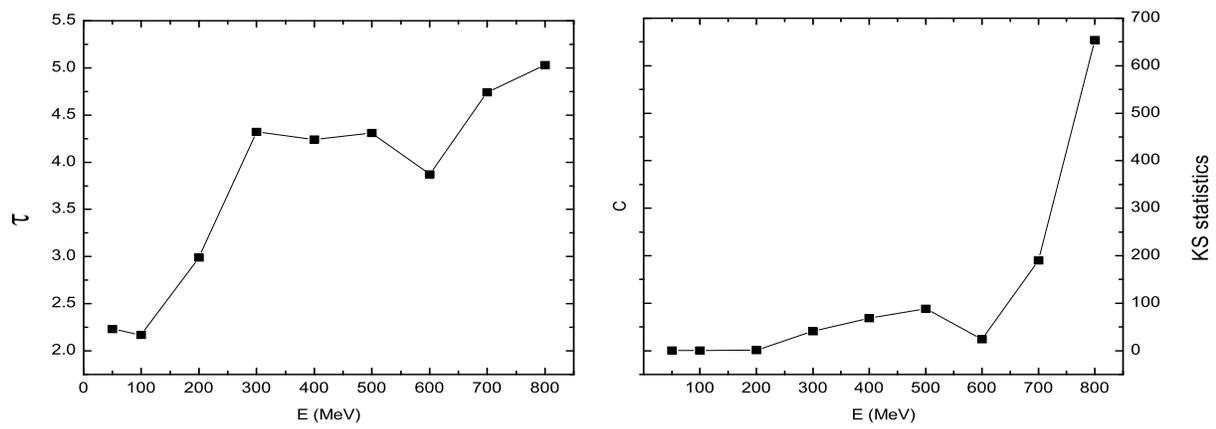,width=20cm,height=12cm,angle=0} }
\caption{The power-law exponents(left) and the KS statistic(right)
for the colliding system as a function of incident energies.}
 \label{Fig9}
\end{figure}

\begin{figure}[htpb]
\centerline{ \epsfig{file=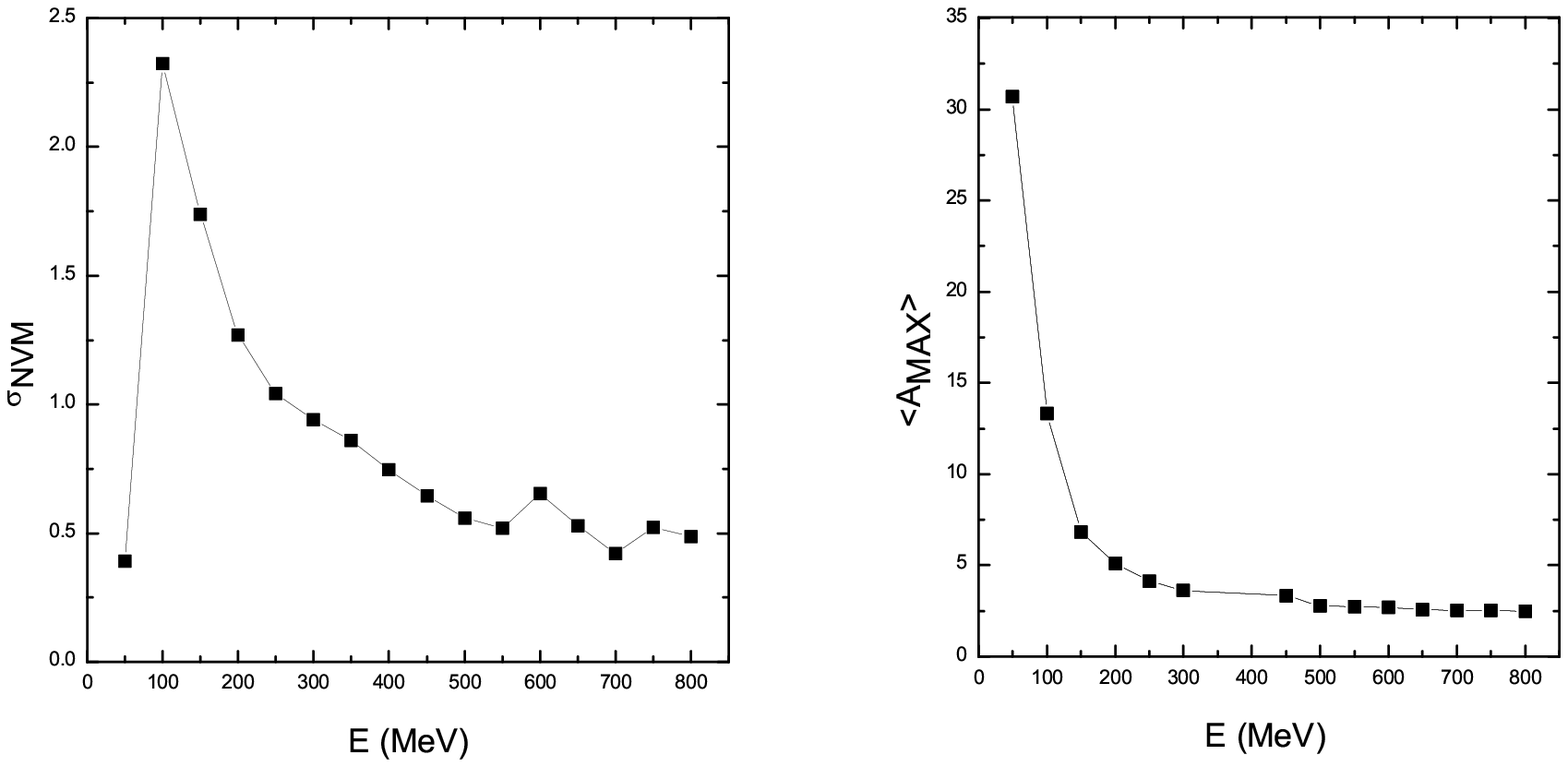,width=22cm,height=16cm,angle=0} }
\caption{The changes of the NVM(left) and the average mass of the largest clusters(right) as a function of incident energy.}
 \label{Fig9}
\end{figure}

\end {document}